\documentclass[english,prd,superscriptaddress,nofootinbib,preprintnumbers,showpacs,preprint]{revtex4}
\usepackage[latin1]{inputenc}
\usepackage{graphicx}
\usepackage{amsmath,amsthm,amssymb}
\usepackage{bm}

\usepackage{amsfonts}
\usepackage{dcolumn}

\usepackage{color}

\newcommand{\beqa}{\begin{eqnarray}}
\newcommand{\eeqa}{\end{eqnarray}}
\newcommand{\simg}{\gtrsim}

\usepackage{babel}
\begin{document}

\title{Consistency Relations for Large Field Inflation
}

\author{Takeshi Chiba}
\affiliation{Department of Physics, College of Humanities and Sciences, 
Nihon University, Tokyo 156-8550, Japan}
\author{Kazunori Kohri}
\affiliation{Institute of Particles and Nuclear Studies, KEK, and Sokendai, Tsukuba 305-0801, Japan}

\begin{abstract}
Consistency relations for chaotic inflation with a monomial potential
 and natural inflation and hilltop inflation are given which involve the scalar 
spectral index $n_s$, 
the tensor-to-scalar ratio $r$ and the running of the spectral index 
$\alpha$.  The measurement of  $\alpha$ with $O(10^{-3})$ 
and the improvement in the measurement of $n_s$  could discriminate 
 monomial model from natural/hilltop inflation models. A consistency region for general large field models is 
 also presented. 
\end{abstract}

\date{\today}


\maketitle

\section{Introduction}

The possible detection of the primordial B-mode \cite{bicep2} has changed 
the landscape of models of inflation. The scene has completely changed  
from small inflation models to large field inflation models, although the plot thickens \cite{Seljak}. 
Awaiting for the polarization results by Planck, in the meantime, we may entertain 
the possibility of large field inflation and shall speculate on the way to further 
narrow down the models of inflation. 
Then the analysis would be inevitably model-dependent. 
However, we would like to minimize the dependence on 
model parameters. So, we consider a relation 
which a given (single field) inflation model predicts 
independent of model parameters, in the same spirit as the single-field inflationary consistency relation \cite{ll}. 

\section{Consistency Relations for Large Field Inflation}
\label{sec2}

Large field models of inflation inhabit the region where 
the scalar spectral index is red $n_s<1$  and 
the tensor-to-scalar ratio is relatively large $r\simg 0.1$ 
\cite{dodelson}. 
Chaotic inflation with a monomial potential  \cite{chaotic} and 
natural inflation \cite{natural} are typical examples of (single field) large field inflation. 
So, we attempt to derive consistency relations for these models 
which hold independent of model parameters.
\footnote{A similar attempt was  
made in \cite{creminelli}, but there the relation for chaotic inflation 
was limited to a quadratic potential (or depends 
on the power index) and the relation 
for natural inflation depends on the model parameter.}
We use the units of $M_{\rm pl}=1/\sqrt{8\pi G}=1$.

\subsection{Monomial Potential}

First, we consider chaotic inflation with a monomial potential:
\beqa
V=\lambda \phi^n,
\eeqa
where we assume $\phi>0$ and the power index $n(>0)$ needs not be integer and can be fractional (or real) 
number like $2/3$ as in axion monodromy inflation model \cite{silverstein}. 
In any case, $n$ is a constant and can be written as $n={d\ln V }/{ d\ln\phi}$. 
Differentiating $n$ with respect $\phi$, we have
\beqa
\phi=\frac{VV'}{V'^2-VV''},
\label{phi=}
\eeqa
where $V'=dV/d\phi$, and so on. Further taking the derivative, we obtain
\beqa
VV'^2V''-2V^2V''^2+V^2V'V'''=0 \,.
\label{eq3}
\eeqa
In addition, since we assume $\phi>0$ (and hence $V'>0$), from 
Eq. (\ref{phi=}) we require 
\beqa
V'^2-VV''>0 \,. 
\label{eq4}
\eeqa
In terms of the slow-roll parameters
\beqa
\epsilon\equiv \frac12\left(\frac{V'}{V}\right)^2,  \quad \eta\equiv\frac{V''}{V},  
\quad \xi\equiv \frac{V'V'''}{V^2} \,,
\eeqa
these relations Eq. (\ref{eq3}) and Eq. (\ref{eq4}) can be rewritten as
\beqa
2\epsilon\eta-2\eta^2+\xi=0   \quad{\rm and}  \quad 2\epsilon>\eta \,.
\label{consist1}
\eeqa
Using inflationary observables related to the slow-roll parameters, 
the scalar spectral index $n_s$, 
the tensor-to-scalar ratio $r$ and the running of the spectral index 
$\alpha$
\beqa
n_s-1=-6\epsilon+2\eta,  \quad r=16\epsilon,  \quad 
\alpha=\frac{d n_s}{d\ln k}=16\epsilon\eta-24\epsilon^2-2\xi \, ,
\eeqa
Eq. (\ref{consist1}) become  relations among observables 
\footnote{We note that the prediction of the running might have been
changed if there had been an additional (dynamical) light field during
inflation ~\cite{Kohri:2014jma}.}
\beqa
\alpha=-(1-n_s)^2 +\frac18 r(1-n_s)   \quad{\rm and}  \quad 
1-n_s>\frac18 r \,,
\label{consist:mono}
\eeqa
which we call consistency relations for monomial chaotic inflation which may 
be reminiscent of the consistency relation for a single field inflation \cite{ll}.  
The second inequality implies the red spectrum: $n_s<1$. 
Note that Eq. (\ref{consist:mono}) holds for chaotic inflation with a monomial potential irrespective of the power index $n$.

\subsection{Natural Inflation}

Next, we consider natural inflation 
\beqa
V=V_0\left(1-\cos\left(\frac{\phi}{f}\right)\right),
\eeqa
where we assume $0<\phi<\pi f$ and 
$f$ is the decay constant and $V_0$ is related with the breaking 
scale of the global symmetry for axion. For $f \gg 1$ the potential becomes 
indistinguishable from a quadratic potential. 

$V_0$ can be written as 
\beqa
V_0^2=(V+f^2V'')^2=(fV')^2+(f^2V'')^2, 
\eeqa
and $f$ can be written as $f^2=-V'/V'''$. Hence, using the slow-roll parameters, 
we obtain a relation 
\beqa
4\epsilon^2-4\epsilon\eta+\xi=0.
\eeqa
Moreover, since $V'>0$ and $V'''<0$, $\xi<0$ is required. 
Then, in terms of observables, we obtain  relations
\beqa
\alpha=\frac{1}{32}r^2-\frac14 r(1-n_s)\quad {\rm and}\quad 
\alpha>\frac{3}{32}r^2-\frac12 r(1-n_s)
\label{consist:natural}
\eeqa
which we call  consistency relations for natural inflation. 
Note that 
 Eq. (\ref{consist:natural}) holds for natural inflation  
irrespective of the value of $f$.  Note that the inequality is saturated when $\alpha=-r^2/32$ which corresponds to the relation for a quadratic potential.  
We also note that the second inequality can also be derived from the inequality 
\beqa
r<4(1-n_s) \,, 
 \label{eq:UppTensExtranatural}
\eeqa  
which follows from $\cos\left(\phi/f\right)=\eta/(2\epsilon-\eta)<1$.

\subsection{Extra Natural Inflation}

The potential of extranatural inflation~\cite{extranatural} is given
by
\begin{eqnarray}
  \label{eq:Vextra}
V(\phi)=V_0\left[
1-\sum_{n=1}^{\infty}\frac{\cos\left(n\frac{\phi}{f}\right)}{n^5}
\right].
\end{eqnarray}
For simplicity,  following \cite{lim},  we neglect  the higher
$n$-terms for $n \ge 2$ to calculate $V, V'$ and $V''$ for both $\epsilon $ and $\eta$ 
since they are suppressed by $1/n^5, 1/n^4$ or $1/n^3$,  but we include higher order terms to calculate $V'''$ (and higher derivatives).  Then $\xi$ is given  approximately by~\cite{lim},
\begin{eqnarray}
  \label{eq:xiextranatural}
  \displaystyle{
  \xi = 
  \frac{
  \left[
      \ln \left(\frac{\phi}{f} \right) - 1
      \right]
\left(\frac{\phi}{f} \right)
    \sin \left(\frac{\phi}{f} \right)
  }
  {
  f^4 \left[ 1 - \cos\left(\frac{\phi}{f} \right) \right]^2
      }
  },
\end{eqnarray}
where
\begin{eqnarray}
  \label{eq:cosine}
  \cos \left(\frac{\phi}{f} \right) = \frac{\eta}{2 \epsilon - \eta}\,,
\end{eqnarray}
and $ \cos\left(\frac{\phi}{f} \right) < 1$  gives the same condition
as ~ (\ref{eq:UppTensExtranatural}) under this approximation. 
{}From Eq. (\ref{eq:xiextranatural}) and Eq. (\ref{eq:cosine}) together with $f^{-2}=2(\epsilon-\eta)$, 
$\xi$ is written as a function of $\epsilon$ and $\eta$, and hence we obtain a relation 
among $n_s, r$ and $\alpha$ which is too complicated to show here. 
Note that the prediction of $r = 16
\epsilon$ could roughly have a 10~$\%$ error at most because $|\Delta
r/r| = |\Delta \epsilon / \epsilon| \sim 2 |\Delta V'/V' | \sim
\Sigma_{n=2}^{\infty} \frac{2}{n^4} \sin(n\phi/f)\ll 1/2^3$. The
validity of this approximation was checked in detail by
Ref.\cite{lim}.


\subsection{Hilltop Inflation}

We can also derive a consistency relation for hilltop \cite{hilltop} 
(or symmetry breaking \cite{kl}) inflation 
\beqa
V(\phi)=\frac{\lambda}{4}\left(\phi^2-v^2\right)^2 \,.
\eeqa
For $\phi\gg v$, the potential becomes a quartic potential. 
A simple calculation gives
\beqa
3\epsilon^2-3\epsilon\eta+\xi=0.
\eeqa
Moreover, since $9V'V'''-6V''^2=-6\lambda^2(3\phi^2+v^2)v^2<0$, 
we have an inequality
\beqa
\xi-\frac23 \eta^2<0 \,. 
\eeqa
In terms of $\alpha,r$ and $n_s$,  consistency relations become 
\beqa
\alpha=\frac{3}{64}r^2-\frac{5}{16}r(1-n_s) \quad {\rm and}\quad 
\alpha>-\frac13(1-n_s)^2+\frac{3}{64}r^2-\frac{1}{4}r(1-n_s)\, .
\label{consist:hilltop}
\eeqa
Note that the inequality is saturated at $\alpha=-(3/256)r^2$ which 
precisely corresponds to the  relation for a 
quartic potential. 

\begin{figure}
\includegraphics[height=3.3in,width=5.0in]{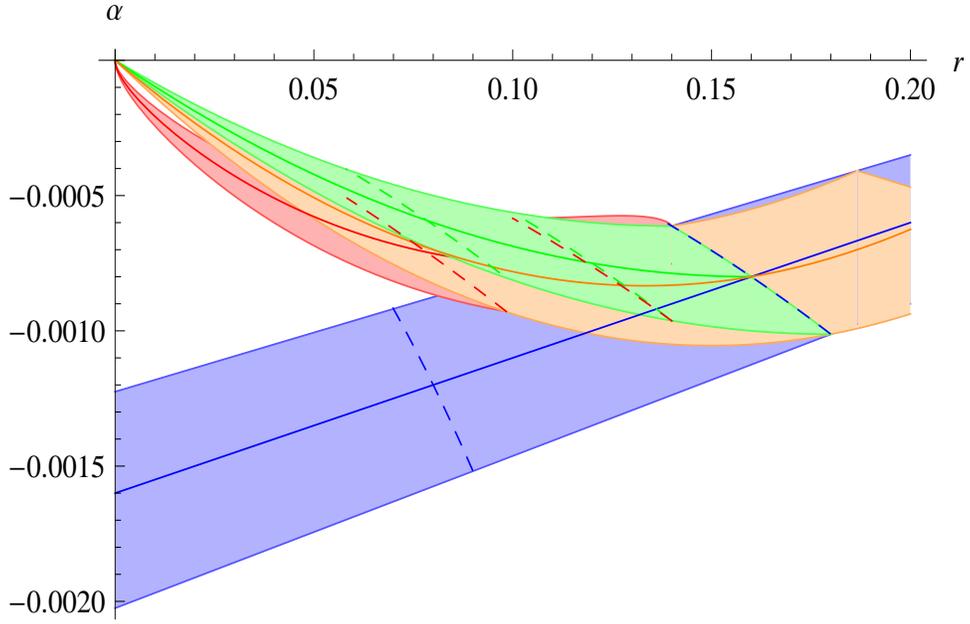}
\caption{\label{fig1}
Consistency relations for $0.955<n_s<0.965$ in $(r,\alpha)$ plane. 
The shaded regions (blue, green, red, orange) are the relations 
for monomial, natural, extranatural, symmetry breaking potential, respectively. 
For each region, the upper (lower) curve is for $n_s=0.965 (0.955)$. 
The middle solid curves are for $n_s=0.96$.  Blue dashed curved are for 
 $n=2/3, 2$ from left to right, and green or red dashed curves are 
for $f=7, 10$ from left to right. }
\end{figure}

\begin{figure}
\includegraphics[height=3.3in,width=5.0in]{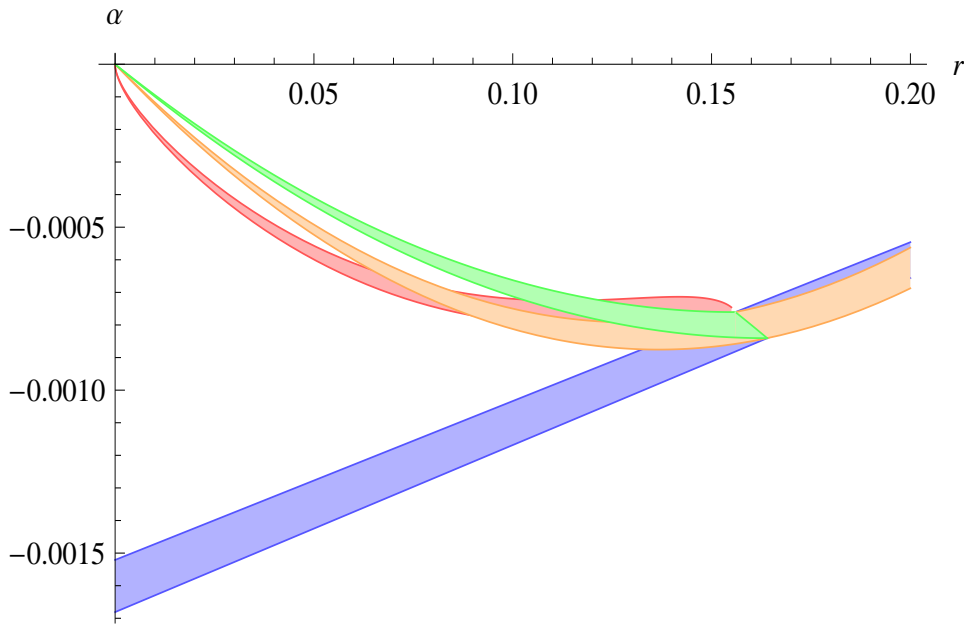}
\caption{\label{fig1b}
Same as Fig. \ref{fig1} but for $n_s=0.96\pm 0.001$. }
\end{figure}

In Fig. \ref{fig1}, we show these relations in $(r,\alpha)$ plane for $0.955<n_s<0.965$ 
which should be possible by measurements by Planck~\cite{planck}.  
The shaded regions (blue, green, red, orange) are the relations 
for monomial potential, natural, extranatural, symmetry breaking potential, respectively.  
For each region, the upper (lower) curve is for $n_s=0.965 (0.955)$. 
The middle solid curves are for $n_s=0.96$.  
 Blue dashed curved are for 
 $n=2/3, 2$ from left to right, and green or red dashed curves are 
for $f=7, 10$ from left to right, although for $f=10$ green dashed curve almost coincides with red dashed curve. 
In Fig. \ref{fig1b}, we also show the relations for for $n_s=0.96\pm 0.001$ 
which might be possible by future observations of the fluctuations of the 21 cm line of 
neutral hydrogen \cite{kohri}. 
 
The current constraint on $\alpha$ from Planck is $\alpha=-0.019\pm 0.010$ \cite{PlanckXVI}. 
The measurement of $\alpha$ with the precision of $O(10^{-3})$, which 
would be possible \cite{kohri} by future observations of the 21 cm line by SKA 
\cite{ska} or by Omniscope 
\cite{omniscope}, 
could discriminate chaotic inflation with a monomial model from 
natural/extranatural/hilltop models. 
Further,  the measurement of $\alpha$ with a precision of $O(10^{-4})$, which
would be possible \cite{kohri} by measurements by CMBPol \cite{cmbpol} combined 
with Omniscope \cite{omniscope} , could discriminate  natural inflation from hilltop inflation. 


\subsection{More General Large Field Models}
\begin{figure}
\includegraphics[height=3.3in,width=5.0in]{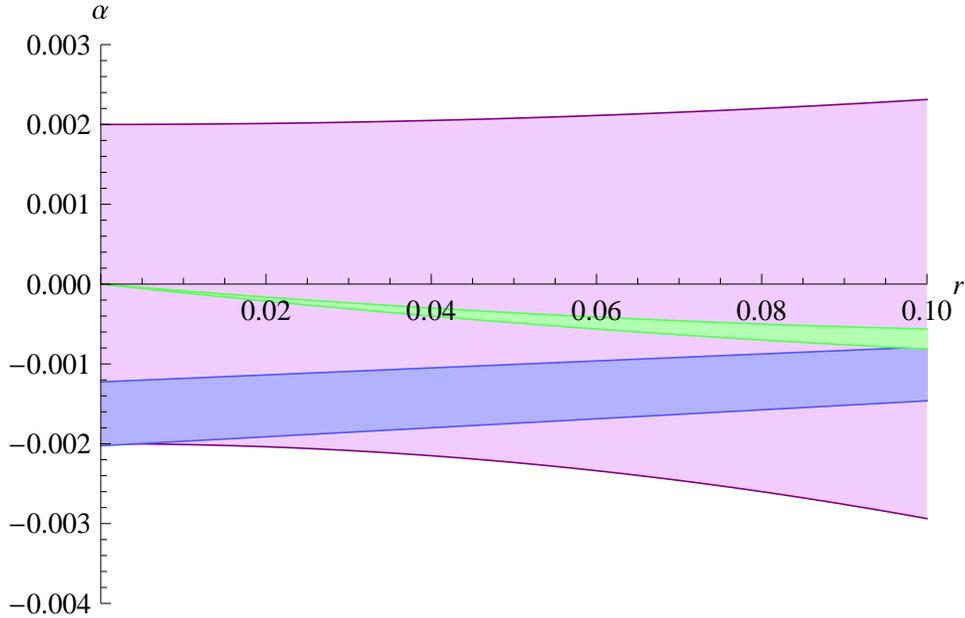}
\caption{\label{fig2}
Consistency region for general large field models for $|\xi|<10^{-3}$. 
The consistency relations for monomial and natural inflation in Fig. \ref{fig1} are also shown. }
\end{figure}

For more general models,  firstly we need to define the large field model. 
Following \cite{dodelson}, we define the large field model by
\beqa
0 < \eta < 2\epsilon,
\label{large}
\eeqa
where the first inequality follows from the convexity of $V$: $V''>0$ \footnote{Therefore, a monomial $\phi^n$ with $n<1$ is no longer a large field model, according to this definition. } 
and the second inequality 
from the exponential function (power-law inflation). 
In this case, $\alpha$ is limited by
\beqa
-\frac{3}{32} r^2-2\xi < \alpha <  \frac{1}{32}r^2 -2\xi.
\eeqa
The inequality involves an unknown parameter $\xi$. However, since $\xi$ is the second order 
slow-roll parameter, it may be at most of $O(N^{-2})\sim O(10^{-3})$, where $N \sim 50\sim 60$ is 
the e-folding number during inflation. 
Therefore, if we vary $\xi $ from $-10^{-3}$ to $10^3$, the region bounded by
\beqa
-\frac{3}{32} r^2-2\times 10^{-3} < \alpha <  \frac{1}{32}r^2 +2 \times 10^3,
\label{inequality}
\eeqa
is the allowed region for general large field models defined by Eq. (\ref{large}). The region is shown 
in Fig. \ref{fig2} together with the consistency relations for monomial and natural  
inflation shown in Fig. \ref{fig1}. 
In any case, the measurement of $\alpha$ with the precision of $O(10^{-3})$ is required to 
probe the region. Conversely,  the measurement of 
$|\alpha| >3\times 10^{-3}$ would refute 
the large field models defined by Eq. (\ref{large}). 

%

\section{Summary} 
\label{secsum}

We have provided consistency relations for chaotic inflation with a monomial potential  
Eq. (\ref{consist:mono}), for natural inflation Eq. (\ref{consist:natural}) and for 
hilltop inflation Eq. (\ref{consist:hilltop}) which relate $n_s, r$ and $\alpha$.  
We have also given an inequality Eq. (\ref{inequality}) 
for large field models defined by  Eq. (\ref{large}).  We find that the running of 
the spectral index $\alpha$ as well as the tensor-to-scalar ratio $r$ is the key observables 
to discriminate monomial models from natural/extranatural inflation models. 
We should emphasize that  $n_s$ and $r$ without using $\alpha$ monomial models cannot be discriminate from natural/extranatural inflation unless we assume the  power index 
of monomial potential and the e-folding number $N$. 
Even for smaller $r$, $ (r,n_s)$ of natural inflation with larger $N$ can overlap with monomial 
with lower $N$.  We stress that $N$ is not a measurable quantity. 

It would be interesting to extend such relations to other large field models, 
such as polynomial models, but that would involve the running of $\alpha$. It would also be interesting to investigate inflation models 
with non-canonical kinetic terms. 
We hope that our consistency relations would help to pin down the inflation model.

\section*{ACKNOWLEDGEMENTS}

We would like to thank T.Suyama, M.Yamaguchi, J.Yokoyama, S.Yokoyama 
and D.Yamauchi for useful comments. 
This work is supported by the Grant-in-Aid for Scientific Research
from JSPS (Nos.\,24540287 (TC), 23540327 and 26105520 (KK)), and in part
by Nihon University (TC), and by the Center for the Promotion of
Integrated Science (CPIS) of Sokendai 1HB5804100 (KK).



\begin{thebibliography}{10}

  
\bibitem{bicep2}
P.~A.~R.~Ade {\it et al.}  [BICEP2 Collaboration],
  Phys.\ Rev.\ Lett.\  {\bf 112}, 241101 (2014)
  [arXiv:1403.3985 [astro-ph.CO]].
  
\bibitem{Seljak} 
M.~J.~Mortonson and U.~Seljak,
arXiv:1405.5857 [astro-ph.CO]; 
R.~Flauger, J.~C.~Hill and D.~N.~Spergel,
arXiv:1405.7351 [astro-ph.CO].


\bibitem{ll}
A.~R.~Liddle and D.~H.~Lyth,
  Phys.\ Lett.\  B {\bf 291}, 391 (1992)
  [arXiv:astro-ph/9208007].

\bibitem{creminelli}
P.~Creminelli, D.~Lopez Nacir, M.~Simonovic, G.~Trevisan and M.~Zaldarriaga,
  arXiv:1404.1065 [astro-ph.CO].

\bibitem{dodelson}
S.~Dodelson, W.~H.~Kinney and E.~W.~Kolb,
  Phys.\ Rev.\ D {\bf 56}, 3207 (1997)
  [astro-ph/9702166].
  
\bibitem{chaotic}
A.~D.~Linde,
  Phys.\ Lett.\ B {\bf 129}, 177 (1983).
  
\bibitem{natural}
 K.~Freese, J.~A.~Frieman and A.~V.~Olinto,
  Phys.\ Rev.\ Lett.\  {\bf 65}, 3233 (1990).

\bibitem{silverstein}
E.~Silverstein and A.~Westphal,
  Phys.\ Rev.\ D {\bf 78}, 106003 (2008)
  [arXiv:0803.3085 [hep-th]].


\bibitem{Kohri:2014jma}
  K.~Kohri and T.~Matsuda,
  arXiv:1405.6769 [astro-ph.CO].

\bibitem{extranatural}
N.~Arkani-Hamed, H.~-C.~Cheng, P.~Creminelli and L.~Randall,
  Phys.\ Rev.\ Lett.\  {\bf 90}, 221302 (2003)
  [hep-th/0301218].

\bibitem{lim}
 K.~Kohri, C.~S.~Lim and C.~-M.~Lin,
  arXiv:1405.0772 [hep-ph].


\bibitem{hilltop}
L.~Boubekeur and D.~.H.~Lyth,
  JCAP {\bf 0507}, 010 (2005)
  [hep-ph/0502047].
  
\bibitem{kl}
A.~D.~Linde,
  Phys.\ Lett.\ B {\bf 132}, 317 (1983); 
  

\bibitem{planck}
J.~Tauber {\it et al.}  [Planck Collaboration],
  astro-ph/0604069.

\bibitem{kohri}
K.~Kohri, Y.~Oyama, T.~Sekiguchi and T.~Takahashi,
  JCAP {\bf 1310}, 065 (2013)
  [arXiv:1303.1688 [astro-ph.CO]].
  
\bibitem{PlanckXVI}
 P.~A.~R.~Ade {\it et al.}  [Planck Collaboration],
  arXiv:1303.5076 [astro-ph.CO].

\bibitem{ska}
 	http://www.skatelescope.org/;

 G.~Mellema, L.~V.~E.~Koopmans, F.~A.~Abdalla, G.~Bernardi, B.~Ciardi, S.~Daiboo, A.~G.~de Bruyn and K.~K.~Datta {\it et al.},
  Exper.\ Astron.\  {\bf 36}, 235 (2013)
  [arXiv:1210.0197 [astro-ph.CO]].

\bibitem{omniscope}
 M.~Tegmark and M.~Zaldarriaga,
  Phys.\ Rev.\ D {\bf 82}, 103501 (2010)
  [arXiv:0909.0001 [astro-ph.CO]].

\bibitem{cmbpol}
D.~Baumann {\it et al.}  [CMBPol Study Team Collaboration],
  AIP Conf.\ Proc.\  {\bf 1141}, 10 (2009)
  [arXiv:0811.3919 [astro-ph]].


\end{thebibliography}
\end{document}